\documentclass{siamltex}

\usepackage{epsfig}
\usepackage{amsmath}
\usepackage{delarray}
\usepackage{citesort}

\newcommand{\CC}{{\cal C}}
\newcommand{\Imm}{\Phi}
\newcommand{\ALG}{{\cal B}}
\newcommand{\BAL}{{\rm BAL}}
\newcommand{\OPT}{{\cal A}}
\newcommand{\DA}{{\rm DA}}
\newcommand{\cS}{{\cal S}}
\newcommand{\rss}{{r_{\rm s}^*}}
\newtheorem{fact}[theorem]{Fact}

\newcommand{\why}[0]{In Equation~(\ref{def_ratio}), we use
$\frac{\OPT(\vec{e})}{\ALG(\vec{e})}$ instead
$\frac{\ALG(\vec{e})}{\OPT(\vec{e})}$ so that the competitive ratios
of different online algorithms are greater than 1 and therefore are
easier to distinguish visually in Figures \ref{fig:buy:tsmcc} and
\ref{fig:buy:acerc}.  In contrast, in Equation~(\ref{eqn_payoff}), we
choose $\frac{\ALG_i(\vec{\sigma}_j)}{\OPT(\vec{\sigma}_j)}$ instead
of $\frac{\OPT(\vec{\sigma}_j)}{\ALG_i(\vec{\sigma}_j)}$ in order to
simplify the linear algebra involved.}

\begin{document}

\title{Optimal Buy-and-Hold Strategies for Financial Markets with
Bounded Daily Returns\thanks{A preliminary version appeared in
Proceedings of the 31st Annual {ACM} Symposium on Theory of Computing,
1999, pages 119--128.}}

\author{Gen-Huey Chen\thanks{Department of Computer Science and
Information Engineering, National Taiwan University, Taipei, Taiwan,
ROC.}
\and Ming-Yang Kao\thanks{Department of Computer Science, Yale
University, New Haven, CT 06520, USA, kao-ming-yang@cs.yale.edu.
Research supported in part by NSF Grant CCR-9531028.}
\and Yuh-Dauh Lyuu\thanks{Department of Computer Science and
Information Engineering, National Taiwan University, Taipei, Taiwan,
ROC. Research supported in part
by NSC Grant 87-2416-H-002-031.}
\and Hsing-Kuo Wong\footnotemark[4]}

\maketitle

\begin{abstract} 
In the context of investment analysis, we formulate an abstract online
computing problem called a {\it planning game} and develop general
tools for solving such a game.  We then use the tools to investigate a
practical {\it buy-and-hold trading} problem faced by long-term
investors in stocks.  We obtain the unique optimal static online
algorithm for the problem and determine its exact competitive ratio.
We also compare this algorithm with the popular dollar averaging
strategy using actual market data.
\end{abstract}



\pagestyle{myheadings}
\markboth{\sc chen, kao, lyuu, and wong}{\sc optimal buy-and-hold
strategies}

\section{Introduction}\label{sec:intro} In an {\it online} problem, an
{\it online} algorithm $\ALG$ is given one input at a time from a
sequence of inputs. $\ALG$ takes an action on each input before seeing
any remaining input.  In contrast, an {\it off-line} algorithm sees
the entire input sequence before it takes any action.  Each action
yields a positive {\it accumulation}.  Let $E$ denote the set of all
admissible input sequences.  Let $\CC(\vec{e})$ denote the (expected)
total accumulation of an online or offline algorithm $\CC$ on $\vec{e}
\in E$.  Let $\OPT$ denote the {\it optimal} offline algorithm, i.e.,
one that produces the largest total accumulation on each admissible input
sequence.  In competitive analysis \cite{sletar85,yao80a,BoEl98},
$\ALG$'s performance is measured by its {\it competitive ratio}
\begin{eqnarray}
\Upsilon_\ALG & = & \sup_{\vec{e} \in E}\frac{\OPT(\vec{e})}{\ALG(\vec{e})}.
\label{def_ratio}
\end{eqnarray}
The {\it online player} seeks
to minimize this ratio by choosing a suitable $\ALG$ while the {\it
adversary} attempts to maximize it by picking $\vec{e}$ after
examining $\ALG$.  This paper assumes that the adversary is {\it
oblivious}, i.e., it fixes the input sequence before $\ALG$ performs
any computation such as generating random bits.

A {\em planning game} is an abstract online problem where the length
of the input sequence is fixed and known a priori to $\ALG$.  This
time horizon feature captures many important online problems including
those for portfolio rebalancing \cite{cover91,CovOrd96,OrdCov96},
asset trading \cite{Bin97,ccekl95,efkt92,ElFiKT97}, secretary
selection \cite{AjMeWa95,CMRS64,Freema83,kaot97.soda}, and bipartite
matching \cite{kaot90,kvv90}.  A planning game is {\it finite} if the
numbers of admissible sequences of actions and inputs are both finite;
otherwise, it is {\it infinite}.  A finite planning game corresponds
to a linear programming problem, where an optimal randomized online
algorithm corresponds to an optimal feasible solution.  Consequently,
we can show that the smallest competitive ratio of any randomized
online algorithm for such a game is the reciprocal of the value of the
game as a zero-sum two-person game.

In this general optimization framework, we investigate the {\it
buy-and-hold trading problem} defined as follows.  An investor starts
with some capital, which is normalized to $1$ dollar, and trades it
for a certain security over $n$ days, which is referred to as the {\em
investment horizon}.  To avoid triviality, we assume $n \geq 2$.  On
each day, the security has only one {\it exchange rate}, i.e., the
number of shares of the security which one unit of capital can buy.
Upon seeing the exchange rate, the investor executes one transaction
for that day and may trade all or part of the remaining capital. All
the capital must be traded by the $n$-th day, and converting the
acquired security back to capital is prohibited.  The total {\it
accumulation} of the investor is the number of shares of the security
she accumulates at the end of the investment horizon.  Note that the
competitive ratio between the adversary's and the investor's
accumulations is exactly the competitive ratio between the dollar
values of the accumulations.  This problem is faced by millions of investors
who save for retirement purposes on a long-term basis; for instance, a
widely popular security for today's investors would be a stock index
fund.

We employ the {\it bounded daily return model}, in which the next
day's exchange rate $e'$ depends on the current day's exchange rate
$e$ with $e/\beta \leq e'\leq e\alpha$ for some fixed
$\alpha,\beta>1$.  The values $n$, $\alpha$, and $1/\beta$ are known a
priori to the investor.  We call $\alpha$ and $1/\beta$ the {\it daily
return bounds}.  Figure \ref{tab:buy:1} gives some stock markets which
enforce such ratios through circuit breakers. This model can also be
regarded as an approximation to the geometric Brownian motion model
used extensively in the finance community
\cite{Hull97,neftci,Lyuu99,sab95,wilmotthd,baxterrennie,duffie}.

\begin{figure}
\begin{center}
\begin{tabular}{|c|c|c|}\hline
  Exchange    & \multicolumn{2}{c|} {Circuit Breaker} \\ \cline{2-3}
              & $\alpha^{-1}$ & $\beta$ \\ \hline
  Amsterdam   & $90\%$       & $110\%$ \\ \hline
  Bangkok     & $90\%$       & $110\%$ \\ \hline
  Paris       & $95\%$       & $110\%$ \\ \hline
  Taipei      & $93\%$       & $107\%$ \\ \hline
  Tel-Aviv    & $95\%$       & $110\%$ \\ \hline
  Tokyo       & $95\%$       & $130\%$ \\ \hline
  Vienna      & $95\%$       & $105\%$ \\ \hline
\end{tabular}
\end{center}
\label{tab:buy:1}
\caption{Circuit breaker rules in various exchanges.}
\end{figure}

A {\it static} algorithm is an online algorithm for the buy-and-hold
trading problem such that for $1 \leq i \leq n$, the (expected) amount
of dollars invested by the algorithm on the $i$-th day is the same for
all exchange rate sequences.  A {\it dynamic} algorithm refers to any
online algorithm for the problem, which is not necessarily static.
The {\it static} buy-and-hold trading problem refers to the case of
the problem where the investor can only use a static algorithm.

We prove that the smallest possible competitive ratio for any
randomized or deterministic static algorithm is
$\frac{n\alpha\beta-(n-1)(\alpha+\beta)+(n-2)}{\alpha\beta-1}$.  We
also obtain a deterministic static algorithm with this competitive
ratio, called the {\it balanced strategy}, and prove that it is the
only optimal deterministic static algorithm.  In comparison, the
popular dollar averaging strategy has a strictly greater competitive
ratio and thus is not optimal. The balanced strategy is so simple that
it can be executed even by those who are not mathematically
sophisticated.  Starting with one dollar initially, the algorithm
invests
$\frac{\alpha(\beta-1)}{n\alpha\beta-(n-1)(\alpha+\beta)+(n-2)}$
dollar on the first day,
$\frac{(\alpha-1)\beta}{n\alpha\beta-(n-1)(\alpha+\beta)+(n-2)}$
dollar on the last day, and
$\frac{(\alpha-1)(\beta-1)}{n\alpha\beta-(n-1)(\alpha+\beta)+(n-2)}$
dollar on each of the other days.

Previously, El-Yaniv {\it et al} \cite{efkt92,ElYan98,ElFiKT97} obtained
optimal online algorithms for this unidirectional trading problem
under the assumption that the daily exchange rates, instead of the
daily returns, are between a pair of upper and lower bounds.  Al-Binali
\cite{Bin97} further studied the same setting in a framework of risk
and reward \cite{MW86}. 
Our model and that of El-Yaniv {\it et al} are each formulated for
real but different regulations of stock and foreign currency markets.
A subtle difference between these models is that their model fixes a
upper bound and a lower bound on the daily exchange rates globally for
the entire investment horizon, while our model sets new bounds
dynamically every day. Interestingly, although this difference might
seem minor, they give rise to mathematical results of very distinct
flavors using significantly different techniques.

Section \ref{sct:buy:game} discusses how to compute optimal randomized
online algorithms for finite planning games.
Section~\ref{sct:buy:static} uses the general analysis in
\S\ref{sct:buy:game} to derive the balanced strategy and compare it
with the dollar averaging strategy.  Section~\ref{sct_open} concludes
the paper with some open problems.

\section{General analysis of finite planning games}\label{sct:buy:game}
A finite planning game $G$ can be regarded as a finite zero-sum
two-person game $\Gamma_H(m,n)$ defined as follows.  For any integer
$k > 0$, let $Z_k=\{1,2,\ldots,k\}$.  The maximizing player is the
online player, whose pure strategies are the deterministic online
algorithms $\ALG_i$ of $G$ indexed with $i \in Z_m$.  The minimizing
player is the adversary of $G$, whose pure strategies are the input
sequences $\vec{\sigma}_j$ of $G$ indexed with $j \in Z_n$.  The
payoff matrix\footnote{\why} $H$ of $\Gamma_H(m,n)$ is defined by
\begin{eqnarray}
    H(i,j)=\frac{\ALG_i(\vec{\sigma}_j)}{\OPT(\vec{\sigma}_j)} > 0,
       \quad i\in Z_m\ \mbox{and}\ j\in Z_n.
\label{eqn_payoff}
\end{eqnarray}
Let $\Imm(Z_k)$ be the set of all probability density functions
defined on $Z_k$.  For $k = n$ or $m$, each $h\in\Imm(Z_k)$ is
regarded as a point in the $k$-dimensional Euclidean space and
represents a mixed strategy that applies the $\ell$-th pure strategy
indexed by $Z_k$ with probability $h(\ell)$.  By von Neumann's minimax
theorem \cite{PetZen96},
\begin{eqnarray}\nonumber
\max_{f\in\Imm(Z_m)} \min_{g\in\Imm(Z_n)}
             \sum_{i=1}^m \sum_{j=1}^n f(i)g(j)H(i,j)
&=& \min_{g\in\Imm(Z_n)} \max_{f\in\Imm(Z_m)}
               \sum_{i=1}^m \sum_{j=1}^n f(i)g(j)H(i,j)
\\ \label{eqn:buy:game2}
&=& \max_{f\in\Imm(Z_m)} \min_{j\in Z_n}
               \sum_{i=1}^m f(i)H(i,j)
\\ \nonumber
&=& \min_{g\in\Imm(Z_n)} \max_{i\in Z_m}
               \sum_{j=1}^n g(j)H(i,j),
\end{eqnarray}
which is called the {\it value} $v^*$ of $\Gamma_H(m,n)$.

Let $r^*$ be the smallest possible competitive ratio of any randomized
online algorithm for $G$; i.e.,
\[r^*  =  \min_{f\in\Imm(Z_m)}\max_{j\in Z_n}
       \frac{\OPT(\vec{\sigma}_j)}
            {\sum_{i=1}^m f(i)\ALG_i(\vec{\sigma}_j)}.
\]
A randomized online algorithm is {\it optimal} if its competitive
ratio is $r^*$.

The next theorem relates $G$ and $\Gamma_H(m,n)$.

\begin{theorem}\label{tm:buy:game0}\

\begin{enumerate}
\item \(r^* = \frac{1}{v^*}.\)
\item \label{tm:buy:game0-2}
An optimal mixed strategy of the online player of $\Gamma_H(m,n)$
induces an optimal randomized online algorithm for $G$, and vice
versa.
\end{enumerate}
\end{theorem}
\begin{proof}
This theorem follows from Equation (\ref{eqn:buy:game2}).
\end{proof}

In light of Theorem~\ref{tm:buy:game0}, we use $G$ and $\Gamma_H(m,n)$
interchangeably.  A main purpose of this paper is to derive the exact
value of $r^*$ and an optimal randomized online algorithm for $G$.
To do so by means of Theorem~\ref{tm:buy:game0}, the {\it primal} and
{\it dual} problems of $\Gamma_H(m,n)$ or $G$ are defined as follows:
\begin{center}
\rm
\begin{tabular}{llccll}
&&&&&
\\
Primal: &           & &       & Dual: &
\\
minimize  & $x^T u_m$  & & & maximize &  $y^T u_n$
\\
subject to & $x^T H \ge u_n^T$   & & & subject to & $H y \le u_m$
\\
& $x \ge 0$ &  & & & $y \ge 0$
\\
&&&&&
\end{tabular}
\end{center}
where $u_k$ is the column vector of $k$ copies of 1.  

For each $j \in Z_n$, let $H^j$ denote the $j$-th column of $H$.
Moreover, let $X$ and $Y$ be the sets of feasible solutions to the
primal and dual problems of $G$, respectively.  Let $\bar{X}$ and
$\bar{Y}$ be the sets of optimal feasible solutions to these problems.
Let $X^*$ and $Y^*$ be the sets of optimal mixed strategies of the
online player and the adversary, respectively.

The next lemma is useful for computing an optimal randomized online
algorithm for $G$ and its competitive ratio via linear programming.

\begin{lemma}\label{tm:buy:game1}
\begin{enumerate}
\item  
For all nonzero $x \in X$ and $y \in Y$, $\frac{x}{x^T u_m}$ and
$\frac{y}{y^T u_n}$ are mixed strategies for the online player and
the adversary, respectively.
\item\label{tm:buy:game1_2} 
$\min_{x \in X} x^T u_m =r^*= \max_{y \in Y} y^T u_n$.
\item  \label{tm:buy:game1_3} 
$X^*= \frac{1}{r^*}{\cdot}\bar{X} \neq \emptyset$, and
$Y^*=\frac{1}{r^*}{\cdot}\bar{Y} \neq \emptyset$.
\item For each nonzero $x \in X$, if $j\in Z_n$ satisfies $x^TH^j
=\min_{ \ell\in Z_n} x^TH^\ell$, then $\vec{\sigma}_j$ is a worst-case
input sequence for the online player's mixed strategy $\frac{x}{x^T
u_m}$.
\end{enumerate}
\end{lemma}
\begin{proof}
This lemma follows from Theorem~\ref{tm:buy:game0} and basics of
linear programming \cite{PetZen96}.
\end{proof}

The next fact is useful for analyzing the uniqueness of an optimal
randomized online algorithm for $G$.

\begin{fact}[see \cite{Rag94}]\label{tm:buy:static6}
For any $x \in \bar{X}$ and $y \in \bar{Y}$, $x$ and $y$ are extreme
points of the convex polyhedra $\bar{X}$ and $\bar{Y}$ if and only
if there is a square submatrix $H'=\left(h_{ij}\right)_{i\in I,j\in
J}$ of $H$ for some $I\subseteq Z_m$ and $J\subseteq Z_n$ with the
following properties:
\begin{enumerate} 
\item \label{tm:buy:static6-1}
$H'$ is nonsingular.  
\item \label{tm:buy:static6-2}
$\sum_{i\in I} h_{ij}x_i = 1$ for all $j\in J$.
\item \label{tm:buy:static6-3} 
$\sum_{j\in J} h_{ij}y_j = 1$ for all $i\in I$.
\item \label{tm:buy:static6-4}
For all $i\not\in I$, $x_i=0$.
\item \label{tm:buy:static6-5}
For all $j\not\in J$, $y_j=0$.
\end{enumerate}
\end{fact}

The next theorem combines Lemma \ref{tm:buy:game1} and Fact
\ref{tm:buy:static6} for the case $m = n$.

\begin{theorem}\label{tm:buy:game3}
Assume that $m=n$ and $H^{-1}$ exists.  Let $x = (u_n^TH^{-1})^T$ and
$y = H^{-1}u_n$.  Further assume $x \geq 0$ and $y \geq 0$.  Let
$b=\frac{x}{x^T u_n}$.
\begin{enumerate} 
\item \label{tm:buy:game3-1-1} Then, $x$ and $y$ are optimal feasible
solutions to the primal and dual problems of $G$, respectively.
\item \label{tm:buy:game3-1-2} \(\Upsilon_\ALG=r^*=x^T u_n\), where
$\ALG$ is the randomized online algorithm corresponding to the online
player's mixed strategy $b$; in other words, $\ALG$ is optimal for
$G$.
\item \label{tm:buy:game3-1-3} For all $j=1,\ldots,n$,
$\frac{\OPT(\vec{\sigma}_j)}{\ALG(\vec{\sigma}_j)}=\Upsilon_\ALG$;
i.e., $\ALG$ has the same performance relative to the adversary's on
every input sequence.
\item \label{tm:buy:game3-2} If every component of $x$ and $y$ is
strictly greater than $0$, then $x$ and $y$ are the only optimal
feasible solutions to the primal and dual problems of $G$, and
consequently, $\ALG$ is the only optimal randomized online algorithm.
\end{enumerate} 
\end{theorem}
\begin{proof} 

Statement \ref{tm:buy:game3-1-1}.  By direct verification, $x \in X$
and $y \in Y$.  Then, since $x^Tu_n=(y^Tu_n)^T=y^Tu_n$, by
Lemma~\ref{tm:buy:game1}(\ref{tm:buy:game1_2}) $x \in \bar{X}$ and $y
\in \bar{Y}$.

Statement \ref{tm:buy:game3-1-2}. Note that $x^T u_n = r^*$ by
Statement \ref{tm:buy:game3-1-1} and
Lemma~\ref{tm:buy:game1}(\ref{tm:buy:game1_2}).  Then, by
Statement~\ref{tm:buy:game3-1-1} and
Lemma~\ref{tm:buy:game1}(\ref{tm:buy:game1_3}), $b$ is an optimal
mixed strategy of the online player.  Thus, this statement follows from
Theorem~\ref{tm:buy:game0}.

Statement \ref{tm:buy:game3-1-3}.  As pointed out in Statement
\ref{tm:buy:game3-1-2}, $x^T u_n = r^*$.  By direct evaluation and
Statement \ref{tm:buy:game3-1-2}, $b^T H = \frac{1}{r^*} u_n =
\frac{1}{\Upsilon_\ALG} u_n$.  Then, this statement follows from the
fact that by definition, the $j$-th component of $b^T H$ equals
$\frac{\ALG(\vec{\sigma}_j)}{\OPT(\vec{\sigma}_j)}$.

Statement \ref{tm:buy:game3-2}. 
To prove the uniqueness of $\ALG$, by
Theorem~\ref{tm:buy:game0}(\ref{tm:buy:game0-2}) and
Lemma~\ref{tm:buy:game1}(\ref{tm:buy:game1_3}), it suffices to show
that $\bar{X}$ has a unique element.  By basics of linear programming
\cite{PetZen96}, $\bar{X}$ has only a finite number of extreme points,
and any element in $X$ is a finite convex combination of these extreme
points.  Thus, it suffices to show that $x$ is the only extreme point
of $\bar{X}$ as follows.
Since $H^{-1}$ exists, $x$ and $y$ are extreme points of $\bar{X}$ and
$\bar{Y}$ by Fact~\ref{tm:buy:static6} with $I=J=Z_n$.  On the other
hand, let $z$ be any extreme point of $\bar{X}$. Since $y$ is an
extreme point of $\bar{Y}$, there is a square submatrix
$H'=\left(h_{ij}\right)_{i\in I,j\in J}$ of $H$ such that $z$ and $y$
satisfy the five conditions in Fact \ref{tm:buy:static6}.  Since
$y_j>0$ for $j\in Z_n$, $J=Z_n$ by Condition \ref{tm:buy:static6-5}.
Since $H'$ is square, $I=Z_n$ and $H'=H$.  Then, by Condition
\ref{tm:buy:static6-2}, \(z^T H = u_n^T\).
Since \(x^T H = u_n^T\), we have $z = x$ as desired.
\end{proof}

\section{Optimal static algorithms}
\label{sct:buy:static}

This section applies the general tools in \S\ref{sct:buy:game} to the
static buy-and-hold trading problem to derive the smallest possible
competitive ratio for static algorithms.

\subsection{Notations}
\label{sct:buy:static1}
As specified in \S\ref{sec:intro}, the investor in the buy-and-hold
trading problem is given $\alpha, \beta$, and $n$ prior to an $n$-day
investment horizon.

For $i \in Z_n$, let $e_i$ be the given security's exchange rate on
the $i$-th day of the investment horizon.  Let $e_0$ be the exchange
rate on the $0$-th day, i.e., the day right before the investment
horizon.  Without loss of generality, we normalize $e_0$ to 1 to
simplify the discussion.  An {\it admissible} exchange rate sequence
is any $\vec{e}=\langle e_1,e_2,\ldots,e_n\rangle$ where
$e_i\in[e_{i-1}\beta^{-1},e_{i-1}\alpha]$.  

As in \S\ref{sec:intro}, let $E$ denote the set of all admissible
exchange rate sequences.  Let $\OPT$ denote the optimal offline
trading algorithm.  Let $\ALG$ be the investor's online trading
algorithm.  After the adversary examines $\ALG$ but before the
investor starts executing $\ALG$, the adversary picks and fixes some
$\vec{e} \in E$.  On the $i$-th day for $i \in Z_n$, upon seeing
$e_i$, $\ALG$ decides the amount of remaining capital to be traded for
shares of the security without knowing any future exchange rate, i.e.,
$e_j$ with $j > i$.  Note that $\OPT(\vec{e})=\max_{1\le i \le n}
e_i$, and $\ALG(\vec{e})=\sum_{i=1}^n a_i e_i$, where $a_i$ is the
(expected) amount of dollars invested by $\ALG$ on the $i$-th day and
depends only on the current and past exchange rate
$e_1,e_2,\ldots,e_i$.

For $i \in Z_n$, the algorithm $\cS_i$ which trades the entire initial
capital of one dollar on the $i$-th day is called the {\it trade-once
algorithm on the $i$-th day}. Note that $\cS_i$ is static and
$\cS_i(\vec{e})=e_i$.

Let $\cS$ be a randomized static algorithm. Let $s_i$ be the expected
amount of dollars invested by $\cS$ on the $i$-th day. Note that
$s_i\ge0$ for all $i$ and $\sum_{i=1}^n s_i =1$. Thus, let $\cS'$ be the
deterministic static algorithm that invests $s_i$ on the $i$-the day.
Also, since the amounts $s_1,\ldots,s_n$ define a probability density
function in $\Imm(Z_n)$, let $\cS''$ be the randomized static
algorithm that applies $\cS_i$ with probability $s_i$.

\begin{lemma}\label{lem_equivalent}
$\cS$, $\cS'$, and $\cS''$ are equivalent in the sense that for all
$\vec{e} \in E$, $\cS(\vec{e})=\cS'(\vec{e})=\cS''(\vec{e})$.
\end{lemma}
\begin{proof} Straightforward. \end{proof} 

By Lemma~\ref{lem_equivalent}, we identify $\cS, \cS'$, and $\cS''$. Also,
let $\rss$ be the smallest competitive ratio for the static
algorithms; then by Lemma~\ref{lem_equivalent},
\begin{eqnarray}\label{eqn:buy:static2}
\rss &=& \inf_{f\in\Imm(Z_n)}\sup_{\vec{e}\in E}
      \frac{\OPT(\vec{e})}{\sum_{i=1}^n f(i)\, \cS_i(\vec{e})}.
\end{eqnarray}

\subsection{Reduction to finite games}
\label{sct:buy:static2}
The static buy-and-hold trading problem is an infinite planning game
because the adversary has an infinite number of pure strategies, while
by Lemma~\ref{lem_equivalent} the online player has $n$ pure
strategies $\cS_i$.  In order to use the tools in
\S\ref{sct:buy:game}, we need to reduce the game to a finite one by
eliminating the adversary's dominated pure strategies, i.e.,
non-worst-case exchange rate sequences, so that the remaining exchange
rate sequences are finite in number.

\begin{figure}
     \centering\epsfig{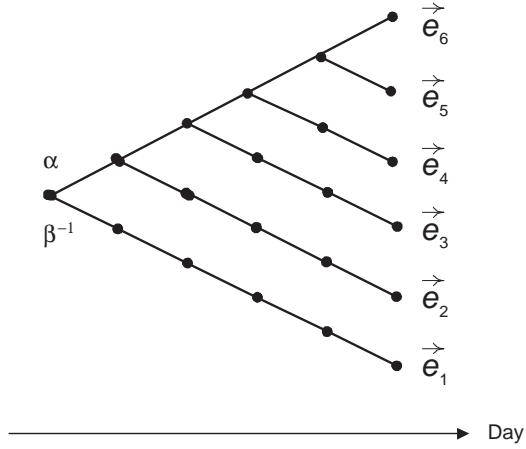}
     \caption{The downturns.}
     \label{fig:buy:1}
\end{figure}

For $j=1,\ldots,n$, let
\[
   \vec{e}_j= \langle\overbrace{\alpha,\alpha^2,\ldots,\alpha^j}^{j},
                   \overbrace{\alpha^j\beta^{-1},\alpha^j\beta^{-2},
           \ldots,\alpha^j\beta^{j-n}}^{n-j}\rangle.
\]
We call these $n$ exchange rate sequences the {\em downturns}; see
Figure~\ref{fig:buy:1} for an illustration.

\begin{lemma}\label{tm:buy_static}
\begin{enumerate}
\item \label{tm:buy:static1} Given a static algorithm $\cS$, each
$\vec{e} \in E$ is dominated by downturn $\vec{e}_j$, i.e.,
$\frac{\OPT(\vec{e})}{\cS(\vec{e})}\leq\frac{\OPT(\vec{e}_j)}{\cS(\vec{e}_j)}$,
where $e_j = \max_{i=1}^n e_i$.
\item\label{tm:buy:static12} The smallest competitive ratio for the
static algorithms is
\[\rss = \inf_{f\in\Imm(Z_n)}\max_{1\le j \le n}
\frac{\OPT(\vec{e}_j)}{\sum_{i=1}^n f(i)\, \cS_i(\vec{e}_j)}.
\]
\item \label{tm:buy:static2} The static buy-and-hold trading problem
can be regarded as a finite zero-sum two-person game $\Gamma_K(n,n)$
with the payoff matrix $K$ defined by
$K(i,j)=\alpha^{i-j}$ if $i\le j$ or $\beta^{j-i}$ if  $i> j$, i.e., 
\[
K=\begin{array}({ccccc})
       1          & \alpha^{-1} & \alpha^{-2} & \cdots  & \alpha^{1-n} \\
       \beta^{-1} & 1           & \alpha^{-1} & \cdots  & \alpha^{2-n} \\
       \beta^{-2} & \beta^{-1}  & 1           & \cdots  & \alpha^{3-n} \\
        \vdots    & \vdots      & \vdots      & \ddots  & \vdots       \\
       \beta^{1-n}& \beta^{2-n} & \beta^{3-n} & \cdots  & 1
   \end{array}.
\]
\end{enumerate}
\end{lemma}
\begin{proof}    
Statement~\ref{tm:buy:static1} follows from the fact that
$\frac{e_j}{e_i}\leq\alpha^{j-i}$ if $i \leq j$ and
$\frac{e_j}{e_i}\leq\beta^{i-j}$ otherwise.
Statement~\ref{tm:buy:static12} follows from Equation
(\ref{eqn:buy:static2}) and Statement \ref{tm:buy:static1}.  For
Statement~\ref{tm:buy:static2}, we let $\cS_i$ be the online player's
$i$-th pure strategy; and let $\vec{e}_j$ be the the adversary's
$j$-th pure strategy. As in \S\ref{sct:buy:game}, the payoff matrix
$K$ is defined by $K(i,j)=\frac{\cS_i(\vec{e}_j)}{\OPT(\vec{e}_j)}$.
The statement then follows from the facts that
$\OPT(\vec{e}_j)=\alpha^j$ and that $\cS_i(\vec{e}_j)=\alpha^i$ if
$i\le j$ or $\alpha^j\beta^{j-i}$ otherwise.
\end{proof}

In light of Lemma~\ref{tm:buy_static}, an optimal mixed strategy of
the online player of $\Gamma_K(n,n)$ corresponds to an optimal static
algorithm. Thus, we next solve $\Gamma_K(n,n)$ to derive an optimal
static algorithm.

\subsection{Deriving an optimal static algorithm}\
\label{sct:buy:static4}

\begin{lemma}\label{tm:buy:static3}
For $n\ge2$, $\det(K)=(1-\alpha^{-1}\beta^{-1})^{n-1} >0$.
\end{lemma}
\begin{proof} 
We use $K_n$ to emphasize the dimension $n$ of $K$.  Let $A_{ij}$ be
the submatrix  of $K_n$ obtained by deleting row $i$ and column $j$.
To expand $\det(K_n)$ along the first row of $K_n$, observe that
$A_{11}=K_{n-1}$.  Furthermore, the first column of $A_{12}$ equals
$\beta^{-1}$ times that of $A_{11}$, while the other columns of
$A_{12}$ equal the corresponding ones of $A_{11}$; thus
$\det(A_{12})=\beta^{-1}\det(A_{11})$.  For $j=3,\ldots,n$,
$\det(A_{1j})=0$ because in $A_{1j}$, the first column equals
$\beta^{-1}$ times the second column. Hence, \(\det(K_n)=\det(A_{11})
-\alpha^{-1}\det(A_{12}) =\det(K_{n-1})
-\alpha^{-1}\beta^{-1}\det(K_{n-1}) =(1-\alpha^{-1}\beta^{-1})
\det(K_{n-1}).\) The lemma immediately follows by induction on $n$.
\end{proof}

Let $b^*$ be the column vector of $n$ components defined by
\begin{gather}
    b_i^* = 
    \begin{cases}
       \frac{\alpha(\beta-1)}{n\alpha\beta-(n-1)(\alpha+\beta)+(n-2)}
          \quad\qquad i=1;\\
       \frac{(\alpha-1)(\beta-1)}{n\alpha\beta-(n-1)(\alpha+\beta)+(n-2)}
          \quad\qquad 1<i<n;\\
       \frac{(\alpha-1)\beta}{n\alpha\beta-(n-1)(\alpha+\beta)+(n-2)}
          \quad\qquad i=n.
    \end{cases}
    \label{eqn:buy:static2x2}
\end{gather}
Since ${b^*} > 0$ and ${b^*}^Tu_n=1$, $b^*$ represents a mixed
strategy of the online player.  So let $\BAL$ denote the static
algorithm which applies $\cS_i$ with probability $b_i^*$; note that by
Lemma~\ref{lem_equivalent}, $\BAL$ is equivalent to the deterministic
static algorithm which invests $b_i^*$ dollars on the $i$-th day.  Let
${c^*}$ be the vector obtained by swapping the first and $n$-th
components of $b^*$.  Similarly, ${c^*} > 0$ and ${c^*}^Tu_n=1$, and
we intend ${c^*}$ to represent a mixed strategy of the adversary.

The next theorem analyzes $\BAL$.  In light of Statement
\ref{thm_balance_property} of the theorem, we call $\BAL$ the {\it
balanced strategy}.

\begin{theorem}\label{thm_optimal}
Let $r=\frac{n\alpha\beta-(n-1)(\alpha+\beta)+(n-2)} {\alpha\beta-1}$.
\begin{enumerate}
\item \label{tmm:buy:static4} 
$\BAL$ is an optimal static algorithm, and \(\Upsilon_\BAL=\rss=r.\)

\item \label{tmm:buy:unique} $\BAL$ is the only optimal static
algorithm subject to the equivalence stated in
Lemma~\ref{lem_equivalent}.

\item \label{thm_balance_property} For
$j=1,\ldots,n,\frac{\OPT(\vec{e}_j)}{\BAL(\vec{e}_j)}=\Upsilon_\BAL$;
in other words, $\BAL$ has the same performance relative to the
adversary's on every downturn.
\end{enumerate}
\end{theorem}
\begin{proof}
Let $\bar{b} = r b^*$, and $\bar{c}= r {c^*}$.  By
Lemma~\ref{tm:buy:static3}, $K^{-1}$ exists.  Below we prove
$\bar{b}^T=u_n^TK^{-1}$ and $\bar{c}=K^{-1}u_n$. Then, Statement
\ref{tmm:buy:static4} follows from
Theorem~\ref{tm:buy:game3}(\ref{tm:buy:game3-1-2}) and the fact that
$\bar{b} \geq 0$, $\bar{c} \geq 0$, and $\bar{b}^T u_n = r$.
Statement \ref{tmm:buy:unique} follows from
Theorem~\ref{tm:buy:game3}(\ref{tm:buy:game3-2}) and the fact that
every component of $\bar{b}$ and $\bar{c}$ is greater than $0$.
Statement \ref{thm_balance_property} follows from
Theorem~\ref{tm:buy:game3}(\ref{tm:buy:game3-1-3})

To prove $\bar{b}^T=u_n^TK^{-1}$ and $\bar{c}=K^{-1}u_n$, observe that
$\bar{c}$ can be obtained by swapping $\alpha$ and $\beta$ in
$\bar{b}$, and the $j$-th column $K^j$ of $K$ can be obtained from the
$j$-th row of $K$ by the same operation.  Therefore,
$\bar{b}^TK=u_n^T$ if and only if $K\bar{c}=u_n$, and we only need to
establish $\bar{b}^TK=u_n^T$.  Since $\bar{b}^TK^1=1$ if and only if
$\bar{b}^TK^n=1$, we only show $\bar{b}^TK^j=1$ for $1\le j <n$ as
follows:
\begin{alignat*}{1}
        \bar{b}^TK^j
        &=\sum_{i=1}^n \bar{b}_i K(i,j) \\
        &=\frac{1}{\alpha\beta-1}\left[
           \alpha(\beta-1) \alpha^{1-j}+
       \sum_{1<i\le j} (\alpha-1)(\beta-1) \alpha^{i-j} +\right.\\
        &\,\qquad\qquad \left.
       \sum_{j<i<n} (\alpha-1)(\beta-1) \beta^{j-i}+
           (\alpha-1)\beta\beta^{j-n}
           \right]\\
        &=\frac{1}{\alpha\beta-1}\left[
          \alpha^{2-j}(\beta-1)+ (\alpha-\alpha^{2-j})(\beta-1)+ \right.\\
        &\,\qquad\qquad \left.
       (\alpha-1)(1-\beta^{j-n+1})+ (\alpha-1)\beta^{j-n+1}
           \right]\\
        &=\frac{1}{\alpha\beta-1}\left(\alpha\beta-1\right)\\
        &=1.
\end{alignat*}
\end{proof}

\subsection{Comparison with the dollar averaging strategy}
\label{sct:buy:static6}
The {\it dollar averaging strategy} ($\DA$) is the static algorithm
which invests an equal amount of capital, i.e., $1/n$ dollars, on each
trading day.  Thus, by Lemma~\ref{lem_equivalent}, $\DA$ is the
uniformly mixed strategy for the online player in the game
$\Gamma_K(n,n)$. By Theorem~\ref{thm_optimal}, $\DA$ is not an optimal
static algorithm, and $\Upsilon_\BAL < \Upsilon_\DA$.  The next lemma
gives a closed-form formula of $\Upsilon_\DA$.  Figure \ref{fig:buy:3}
plots the relationship between $\Upsilon_{\DA}$ and $\Upsilon_{\BAL}$
for $2\le n \le 100$.

\begin{lemma}
\(
\Upsilon_\DA = \max\left\{ \frac{n (1-\alpha^{-1})}{1 - \alpha^{-n}},
\frac{n (1-\beta^{-1})}{1 - \beta^{-n}}\right\}.
\)
\end{lemma}
\begin{proof} 
Let $B_j = \sum_{i=1}^n K(i,j)$.  By Lemma~\ref{tm:buy_static},
$\Upsilon_{\DA}=\max_{1\le j \le n} \frac{n}{B_j}$.  By algebra,
$B_{j+1}-B_j$ is a decreasing function of $j$. Thus, $B_j$ is a
function of $j$ whose minimum occurs at one end of the domain
$\{1,\ldots,n\}$.  The lemma follows from this concavity.
\end{proof}

\newcommand{\buffspace}{\vspace{0.05\textwidth}}

\begin{figure}
\buffspace
\centering\epsfig{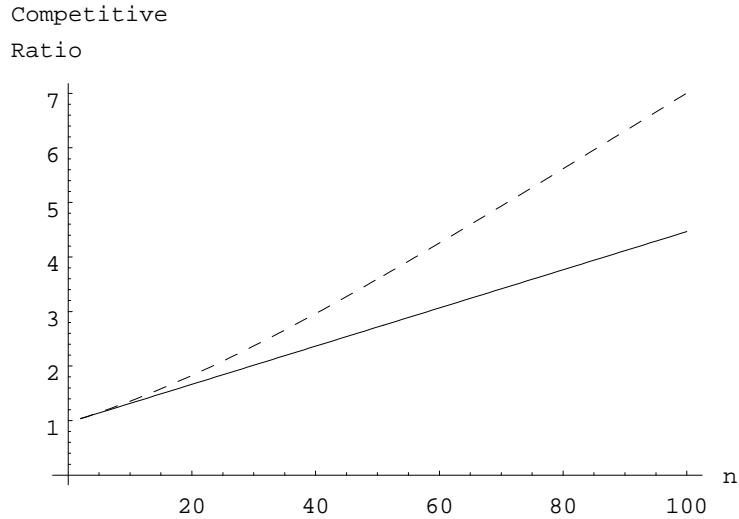}
\buffspace
\caption{The dashed and solid dotted lines denote $\Upsilon_{\DA}$ and
$\Upsilon_{\BAL}$, respectively, with $\alpha=1/0.93$ and
$\beta=1.07$.}
\label{fig:buy:3}
\end{figure}

We have also experimented with $\BAL$ and $\DA$ using Taiwan's market
data. As shown in Figure~\ref{tab:buy:1}, the Taipei Stock Exchange
(TSE) adopts $\alpha=1/0.93$ and
$\beta=1.07$.  We select the Taiwan
Semiconductor Manufacturing Company (TSMC) and Acer Computer Company
(Acer) for experimental analysis.  TSMC is the largest foundry of
wafer manufacturing in the world and is listed on both TSE and the New
York Stock Exchange (NYSE) under the symbol TSM.  Acer is the world's
third largest PC manufacturer as well as the fifth largest mobile PC
manufacturer.

\begin{figure}
\buffspace
\centering\epsfig{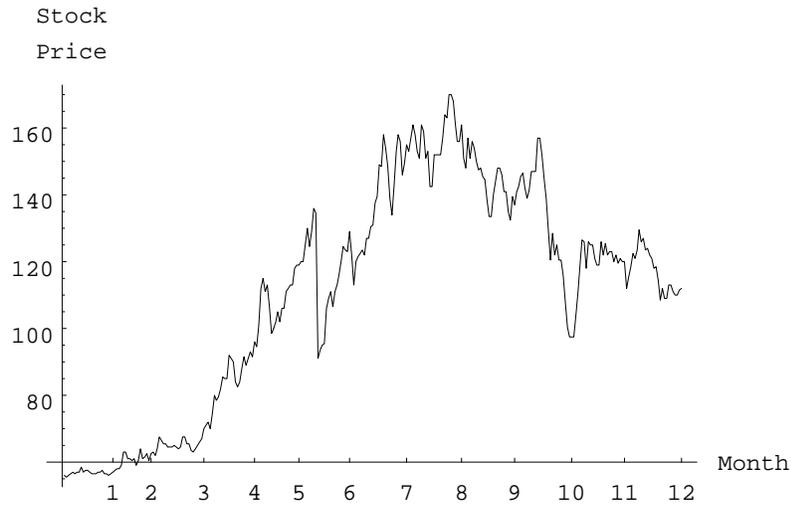}
\buffspace
\caption{TSMC's daily closing stock prices in 1997.}
\label{fig:buy:tsmcp}
\end{figure}

\begin{figure}
\buffspace
\centering\epsfig{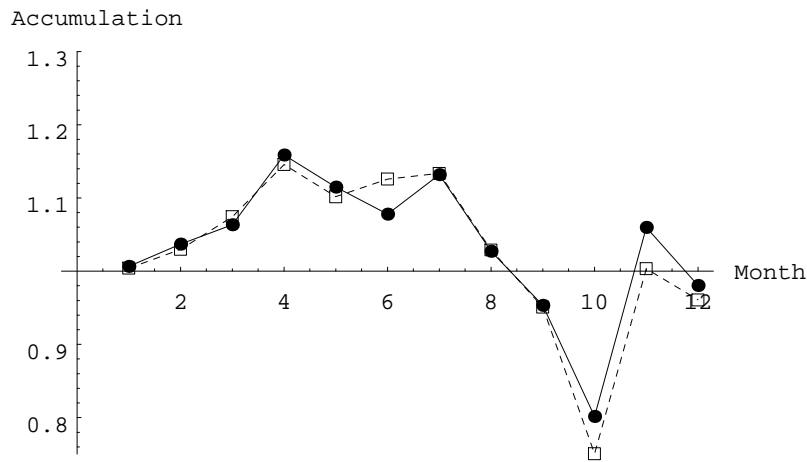}
\buffspace
\caption{Accumulations of $\BAL$ and $\DA$ on TSMC.}
\label{fig:buy:tsmcr}
\end{figure}

\begin{figure}
\buffspace
\centering\epsfig{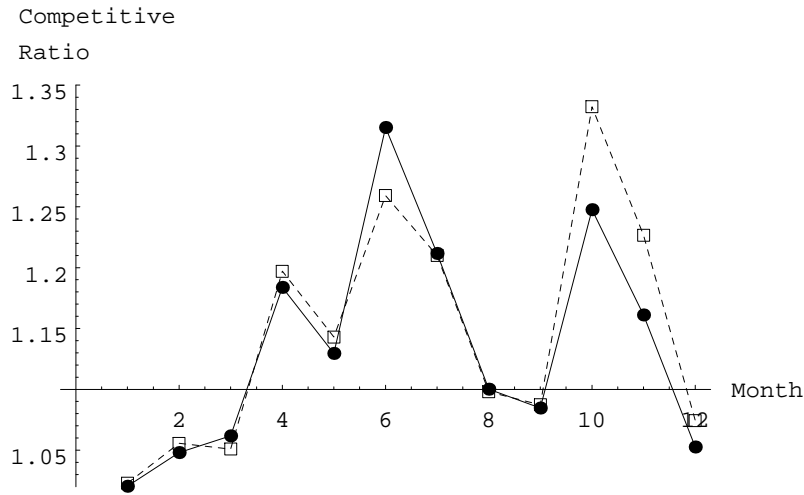}
\buffspace
\caption{Realized competitive ratios of $\BAL$ and $\DA$ on TSMC.}
\label{fig:buy:tsmcc}
\end{figure}

Figure \ref{fig:buy:tsmcp} shows the daily closing prices of TSMC in
1997. All stock prices are quoted in the New Taiwan dollar (NT
dollar).  One investment plan is executed each month. Each plan buys
shares of TSMC with an initial capital of one NT dollar as in
\S\ref{sct:buy:static}; however, the exchange rate of a day is the
reciprocal of that day's share price without an initial normalization
to one. A {\it monthly} accumulation is the total number of shares
acquired over a month.  For ease of comparison, a monthly accumulation
is expressed in NT dollar by converting the acquired shares into NT
dollars at the price of the last trading day of each month.  Figure
\ref{fig:buy:tsmcr} shows the monthly accumulations of $\BAL$ and
$\DA$ on TSMC for each month of 1997.  Notice that $\BAL$ and $\DA$
are money-making except in September, October, and December.  Figure
\ref{fig:buy:tsmcc} shows the {\it realized} competitive ratios of
$\BAL$ and $\DA$, which are the performance ratios as defined in
Equation (\ref{def_ratio}) but with $\vec{e}$ set to the actual
exchange rate sequences.  Note that for all twelve months, these
ratios are less than $1.35$. For visual clarity, we join the monthly
accumulations and competitive ratios by line segments, and use the
solid and dotted lines to denote the graphs of $\BAL$ and $\DA$,
respectively.  Observe that overall, $\BAL$ outperforms $\DA$.

\begin{figure}
\buffspace
\centering\epsfig{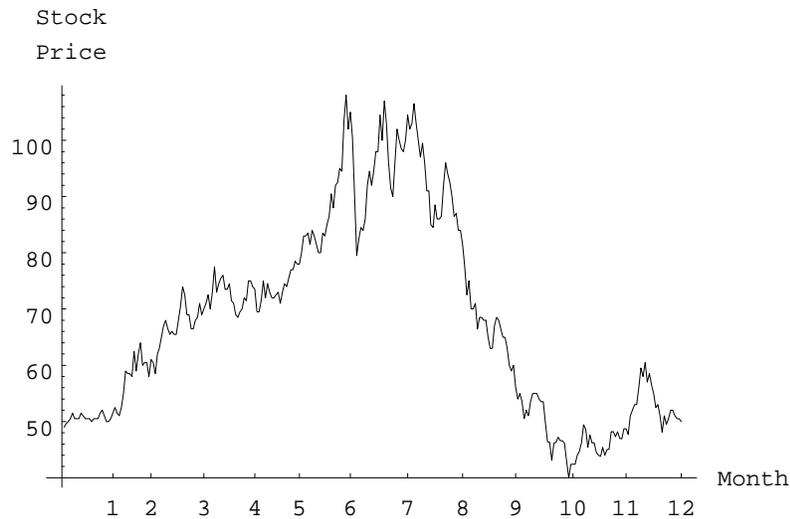}
\buffspace
\caption{Acer's daily closing stock prices in 1997.}
\label{fig:buy:acerp}
\end{figure}

\begin{figure}
\buffspace
\centering\epsfig{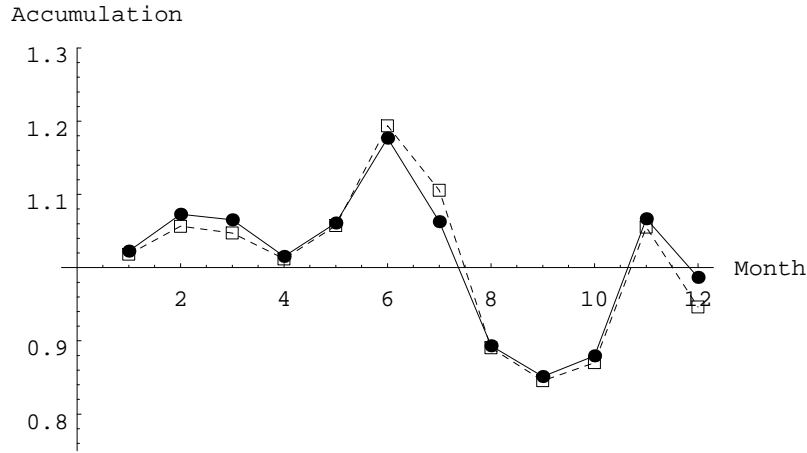}
\buffspace
\caption{Accumulations of $\BAL$ and $\DA$ on Acer.}
\label{fig:buy:acerr}
\end{figure}

\begin{figure}
\buffspace
\centering\epsfig{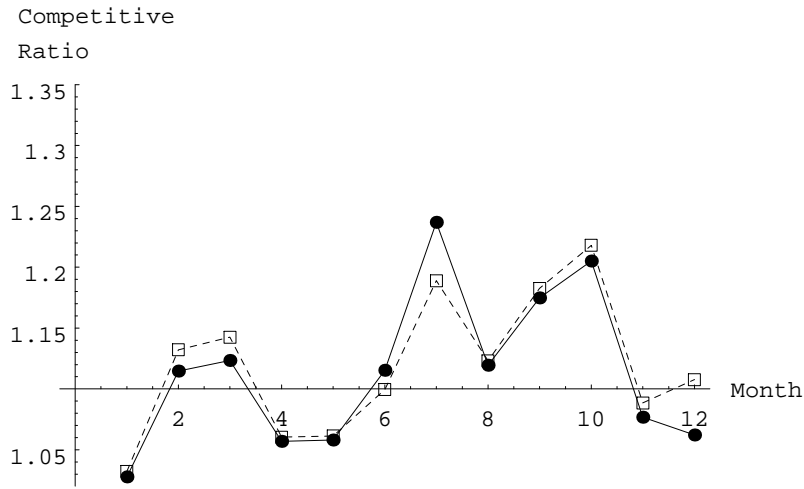}
\buffspace
\caption{Realized competitive ratios of $\BAL$ and $\DA$ on Acer.}
\label{fig:buy:acerc}
\end{figure}

Figure \ref{fig:buy:acerp} shows the daily closing prices of Acer in
1997. Figures \ref{fig:buy:acerr} and \ref{fig:buy:acerc} show the
monthly accumulations and realized competitive ratios of $\BAL$ and $\DA$,
respectively. The experimental results for Acer lead to similar
conclusions to those for TSMC.

\section{Open problems}\label{sct_open}
We have presented the balanced strategy $\BAL$ and proved its unique
optimality among the static algorithms. Furthermore, each of its exact
competitive ratio and daily investment amounts has a closed-form
expression which takes $O(1)$ time to evaluate.  In light of these
results, an immediate open problem is whether there are similar
results for dynamic online trading algorithms.  There are two
orthogonal directions for further research as follows.

One direction is to change the assumption that the time horizon is
fixed and known a priori to $\ALG$. For instance, it would be
meaningful to consider the scenario that there is a cash stream
instead of a one-time capital at the beginning of the investment
horizon. For this scenario, an investor might need to guess when the
cash stream will end.

The other direction is to replace $\alpha$ and $\beta$ with a known
probability distribution of the ratio $\frac{e'}{e}$. This would be an
example of the standard approach in finance of considering the
average-case performance under an assumed probabilistic model.  While
the worst-case approach in computer science is unnecessarily
pessimistic, the average-case approach in finance is overly dependent
on the chosen model. In general. it would be of interest to combine
these two approaches to formulate more informative computational
problems than either approach could.

\section*{Acknowledgments} 
We wish to thank the anonymous referees for very thoughtful comments.
Some of the comments have resulted in open problems in
\S\ref{sct_open}.

\bibliographystyle{siam} \bibliography{all}
\end{document}